\overfullrule=0pt
\input harvmac

\def\a{\alpha}
\def\ab{\overline{\alpha}}
\def\b{\beta}
\def\bb{\overline\b}
\def\g{\gamma}
\def\gb{\overline\g}
\def\d{\delta}
\def\db{\overline\d}
\def\e{\epsilon}
\def\eb{\overline\e}
\def\r{\rho}
\def\rb{\overline\r}
\def\s{\sigma}
\def\sb{\overline\s}
\def\l{\lambda}
\def\lt{\widetilde\l}
\def\lh{\widehat\lambda}
\def\lth{\widehat{\lt}}
\def\o{\omega}
\def\ot{\widetilde\o}
\def\oh{\widehat\o}
\def\oth{\widehat{\ot}}
\def\O{\Omega}
\def\Ob{\overline\O}
\def\Ot{\widetilde\O}
\def\dt{\widetilde d}
\def\dth{\widehat{\dt}}
\def\t{\theta}
\def\tt{\widetilde\t}

\def\F{\Phi}
\def\L{\Lambda}
\def\Lt{\widetilde\L}
\def\D{\Delta}
\def\N{\nabla}
\def\Nb{\overline\nabla}
\def\Nt{\widetilde\N}
\def\Pb{\overline\Pi}
\def\p{\partial}
\def\pb{\overline\partial}
\def\dh{\widehat d}

\def\Ct{\widetilde C}
\def\Qt{\widetilde Q}
\def\St{\widetilde S}
\def\jt{\widetilde j}
\def\Rt{\widetilde R}
\def\Tt{\widetilde T}
\def\Cb{\overline C}
\def\Db{\overline D}
\def\Hb{\overline H}
\def\Ib{\overline I}
\def\Jt{\widetilde J}
\def\Nn{\widetilde N}

\Title{ \vbox{\baselineskip12pt \hbox{IFT-P.033/2006}}}
{\vbox{\centerline{ One-loop Conformal Invariance of the Type II }
\smallskip
\centerline{ Pure Spinor Superstring in a Curved Background}}}

\smallskip
\centerline{Oscar A. Bedoya\foot{e-mail: abedoya@ift.unesp.br}}
\smallskip
\centerline{\it Instituto de F\'{\i}sica Te\'orica, Universidade
Estadual Paulista} \centerline{\it Rua Pamplona 145, 01405-900,
S\~ao Paulo, SP, Brasil}
\bigskip
\centerline{Osvaldo Chand\'{\i}a\foot{e-mail: ochandia@unab.cl}}
\smallskip
\centerline{\it Departamento de Ciencias F\'{\i}sicas, Universidad
Andr\'es Bello} \centerline{\it Rep\'ublica 252, Santiago, Chile}

\bigskip

\noindent We compute the one-loop beta functions for the Type II
superstring using the pure spinor formalism in a generic
supergravity background. It is known that the classical pure
spinor BRST symmetry puts the background fields on-shell. In this paper
we show that the one-loop beta functions vanish as a consequence of
the classical BRST symmetry of the action.

\Date{September 2006}

%\draft

\newsec{Introduction}

It is well known that superstrings are not consistent on any
background field. In the simplest case of a closed bosonic string
coupled to a curved background, the quantum preservation of the Weyl
symmetry implies that the background metric satisfies the Einstein
equations plus $\a'$-corrections. The Weyl symmetry preservation can
be seen as the absence of ultra-violet (UV) divergences in the
quantum effective action \ref\AlvarezGaumeHN{L.~Alvarez-Gaume,
D.~Z.~Freedman and S.~Mukhi, ``The Background Field Method and the
Ultraviolet Structure of the Supersymmetric Nonlinear Sigma Model,''
Annals Phys.\  134 (1981) 85.}.

In the case of superstrings in curved background, the preservation
of Weyl symmetry at the quantum level is much more involved,
although the case where only Neveu Schwarz-Neveu Schwarz
background fields are turned on is very similar to the bosonic
string case \ref\CallanIA{C.~G.~.~Callan, E.~J.~Martinec,
M.~J.~Perry and D.~Friedan, ``Strings in Background Fields,''
Nucl.\ Phys.\ B262 (1985) 593.}. Once we allow the full
supersymmetric multiplet to be turned on, we need a manifestly
supersymmetric space-time sigma model in order to study the
corresponding quantum regime. The Green-Schwarz formalism provides
us with a sigma model action which is manifestly space-time
supersymmetric, nevertheless, one does not manifestly preserve
this symmetry if one tries to quantize.

There exists another formalism for the superstring which does not
suffer of this disadvantage, known as the pure spinor formalism
\ref\BerkovitsFE{N. Berkovits, ``Super-Poincare Covariant
Quantization of the Superstring,'' JHEP 0004 (2000) 018
[arXiv:hep-th/0001035].}.  In this formalism the space-time
supersymmetry is manifest and the quantization is straightforward by
requiring a BRST-like symmetry. We will briefly discuss the basics
of this model in section 2, now let us mention what have been done
to check the consistency of this formalism. The spectrum of the
model is equivalent to the Green-Schwarz spectrum in the light-cone
gauge \ref\BerkovitsNN{N.~Berkovits, ``Cohomology in the Pure Spinor
Formalism for the Superstring,'' JHEP 0009 (2000) 046
[arXiv:hep-th/0006003]\semi N.~Berkovits and O.~Chand\'{\i}a,
``Lorentz Invariance of the Pure Spinor BRST Cohomology for the
Superstring,'' Phys. Lett. B514 (2001) 394 [arXiv:hep-th/0105149].}
and it also allows to find the physical spectrum in a manifestly
super-Poincar\'e covariant manner \ref\BerkovitsQX{N.~Berkovits and
O.~Chand\'{\i}a, ``Massive Superstring Vertex Operator in D = 10
Superspace,'' JHEP 0208 (2002) 040 [arXiv:hep-th/0204121].}. The
pure spinor formalism is suitable to describe strings in background
fields with Ramond-Ramond fields strengths turned on, as it happens
on anti-de Sitter geometries. In this case, it has been checked the
classical \ref\BerkovitsYR{N.~Berkovits and O.~Chand\'{\i}a,
``Superstring Vertex Operators in an AdS$_5\times$S$^5$
Background,'' Nucl. Phys. B596 (2001) 185 [arXiv:hep-th/0009168].}
and the quantum BRST invariance of the model
\ref\BerkovitsXU{N.~Berkovits, ``Quantum Consistency of the
Superstring in AdS$_5\times$S$^5$ Background,'' JHEP 0503 (2005) 041
[arXiv:hep-th/0411170].} as well as its quantum conformal invariance
\ref\ValliloMH{B.~C.~Vallilo, One Loop Conformal Invariance of the
Superstring in an AdS$_5\times$S$^5$ Background,'' JHEP 0212 (2002)
042 [arXiv:hep-th/0210064].} \BerkovitsXU\ (see also
\ref\KlusonWQ{J.~Kluson, Note About Classical Dynamics of Pure
Spinor String on AdS$_5\times$S$^5$ Background,'' hep-th/0603228.}
and \ref\BianchiIM{M.~Bianchi and J.~Kluson, ``Current Algebra of
the Pure Spinor Superstring in AdS$_5\times$S$^5$,'' JHEP 0608
(2006) 030 [arXiv:hep-th/0606188].}).

The superstring in the pure spinor formalism can be coupled to
a generic supergravity background field, as it was shown in
\ref\BerkovitsUE{N.~Berkovits and P.~S.~Howe, ``Ten-dimensional
Supergravity Constraints from the Pure Spinor Formalism for the
Superstring,'' Nucl.\ Phys.\ B635 (2002) 75
[arXiv:hep-th/0112160].}, where a sigma model action was written
for the Heterotic and Type II superstrings.
Here, the classical BRST invariance puts
the background fields on-shell. In the heterotic string case, the
background fields satisfy the ten-dimensional N=1 supergravity
equations plus the super Yang-Mills equations in a curved
background. In the type II case, the background fields satisfy the
ten-dimensional type II supergravity equations. Of course, it could
be very interesting to obtain $\a'$-corrections to these equations
in this formalism by requiring the quantum preservation of some
symmetries of the classical sigma-model action. Since the lowest
order in $\a'$ BRST symmetry puts the background fields on shell,
one expects that the equations of motion for the background fields
derived from the beta function calculation are implied by the BRST
symmetry. This property was verified for the heterotic string, in
whose case, the classical BRST symmetry implies one-loop conformal
invariance \ref\ChandiaHN{O.~Chand\'{\i}a and B.~C.~Vallilo,
``Conformal Invariance of the Pure Spinor Superstring in a Curved
Background,'' JHEP 0404 (2004) 041 [arXiv:hep-th/0401226].}. In this
paper we show that the same property is also valid for the Type II
superstring\foot{There is a more recent development
\ref\BerkovitsNMPS{N. Berkovits, ``Pure Spinor Formalism as an N=2
Topological String'', JHEP 0510 (2005) 089 [arXiv:hep-th/0509120].}
with an even richer world-sheet structure, called non-minimal pure
spinor formalism, which is interpreted as a critical topological
string. Nevertheless, in this paper we restrict to the so called
minimal pure spinor formalism of \BerkovitsFE.}.

In section 2 we review the type II sigma-model construction of
\BerkovitsUE. In section 3 we expand the action by using a covariant
background field expansion. In section 4 we determine the UV
divergent part of the effective action at the one-loop level and
finally, in section 5, we use the expanded action of section 3, to write
from the UV divergent part found in section 4, the beta functions for the
Type II superstring. Then we show that beta functions vanish after
using the constraints on the background fields implied by the classical
BRST invariance of the sigma-model action\foot{In a similar way, using the
he hybrid formalism\ref\BerkovitsHF{N.
 Berkovits, ``Covariant Quantization of the Green-Schwarz Superstring 
in a Calabi-Yau Background'', Nucl. Phys. B431 (1994) 258, 
[arXiv:hep-th/9404162].}, in \ref\Nedel{D. L. Nedel, ``Consistency 
of Superspace Low Energy Equations of Motion of 4D Type II Superstring 
with Type II Sigma Model at Tree-Level'', Phys. Lett. B573 (2003) 217, 
[arXiv: hep-th/0306166].} and \ref\Nedel{Daniel. L. Nedel, ``Superspace 
Type II 4D Supergravity from Type II Superstring, PoS WC2004:020, 2004, 
[arXiv:hep-th/0412152].} were derived the type II 4D supergravity equations 
of motion in superspace by requiring 
superconformal invariance.}.

\newsec{Classical BRST Constraints}

The pure spinor closed string action in flat space-time is defined
by the superspace coordinates $X^m$ with $m = 0, \dots, 9$ and the
conjugate pairs $( p_\a, \t^\a) , (\widetilde{p}_{\ab}, \tt^{\ab})$
with $(\a, \ab) = 1, \dots, 16$. For the type IIA superstring the
spinor indices $\a$ and $ \ab$ have the opposite chirality while for
the type IIB superstring they have the same chirality. In order to
define a conformal invariant system we need to include a pair of
pure spinor ghost variables $(\l^\a, \o_\a)$ and $(\lt^{\ab},
\ot_{\ab})$. These ghosts are constrained to satisfy the pure spinor
conditions $(\l \g^m \l) = (\lt \g^m \lt) = 0$, where $\g^m_{\a\b}$
and $\g^m_{\ab\bb}$ are the $16\times 16$ symmetric ten dimensional
gamma matrices. Because of the pure spinor conditions, $\o$ and $\ot$ are
defined up to $\d\o = (\l\g^m) \L_m$ and $\d\ot = (\lt\g^m) \Lt_m$.
The quantization of the model is performed after the construction of
the BRST-like charges $Q = \oint \l^\a d_\a, \Qt = \oint \lt^{\ab}
\dt_{\ab}$, here $d_\a$ and $\dt_{\ab}$ are the world-sheet
variables corresponding to the $N=2 \,\, D=10$ space-time
supersymmetric derivatives and are supersymmetric combinations of
the space-time superspace coordinates of conformal weights $(1,0)$
and $(0,1)$ respectively. The action in flat space is a free action
involving the above fields, that is

\eqn\actionflat{ S = {1\over{2\pi\a'}} \int d^2 z ~ (\ha \p X^m \pb
X_m + p_\a \pb \t^\a + \widetilde{p}_{\ab} \p \tt^{\ab}) + S_{pure},} where
$S_{pure}$ is the action for the pure spinor ghosts.

In a curved background, the pure spinor sigma model action for the
type II superstring is obtained by adding to the flat action of
\actionflat\ the integrated vertex operator for supergravity
massless states and then covariantizing respect to ten dimensional
$N = 2$ super-reparameterization invariance. The result of doing
this is

\eqn\action{ S = {1\over{2\pi\a'}} \int d^2 z ~ (\ha \Pi^a \Pb^b
\eta_{ab} + \ha \Pi^A \Pb^B B_{BA} + d_\a  \Pb^\a + \dt_{\ab}
\Pi^{\ab} + (\l^\a \o_\b) \Ob_\a{}^\b + (\lt^{\ab} \ot_{\bb})
\Ot_{\ab}{}^{\bb}} $$ + d_\a \dt_{\bb} P^{\a\bb} + (\l^\a \o_\b)
\dt_{\gb} C_\a{}^{\b\gb} + (\lt^{\ab} \ot_{\bb}) d_\g
\Ct_{\ab}{}^{\bb\g} + (\l^\a \o_\b)  (\lt^{\ab} \ot_{\bb})
S_{\a\ab}{}^{\b\bb} )+ S_{pure} + S_{FT},$$ where $\Pi^A = \p Z^M
E_M{}^A, \Pb^A = \pb Z^M E_M{}^A$ with $E_M{}^A$ the supervielbein
and $Z^M$ are the curved superspace coordinates, $B_{BA}$ is the
super two-form potential. The connections appears as $\Ob_\a{}^\b =
\pb Z^M \O_{M\a}{}^\b = \Pb^A \O_{A\a}{}^\b$ and $\Ot_{\ab}{}^{\bb}
= \p Z^M \Ot_{M\ab}{}^{\bb} = \Pi^A \Ot_{A\ab}{}^{\bb}$. They are
independent since the action of \action\ has two independent Lorentz
symmetry transformations. One acts on the $\a$-type indices and
the other acts on the $\ab$-type indices. $S_{pure}$ is the
action for the pure spinor ghosts and is the same as in the flat
space case of \actionflat.

As was shown in \BerkovitsUE, the gravitini and the dilatini
fields  are described by the lowest $\t$-components of the
superfields $C_{\a}{}^{\b\gb}$ and $\Ct_{\ab}{}^{\bb\g}$, while the
Ramond-Ramond field strengths are in the superfield $P^{\a\bb}$. The
dilaton is the theta independent part of the superfield $\Phi$ which
defines the Fradkin-Tseytlin term

\eqn\ft{ S_{FT} = {1\over{2\pi}} \int d^2z ~ r ~ \Phi,} where $r$ is
the world-sheet curvature. Because of the pure spinor constraints, the
superfields in \action\ cannot be arbitrary. In fact, it is
necessary that

\eqn\fields{\O_{A\a}{}^\b = \O_A \d_\a{}^\b + {1\over 4} \O_{Aab}
(\g^{ab})_\a{}^\b,\quad \Ot_{A\ab}{}^{\bb} = \Ot_A \d_{\ab}{}^{\bb}
+ {1\over 4} \Ot_{Aab} (\g^{ab})_{\ab}{}^{\bb},}
$$
C_\a{}^{\b\gb} = C^{\gb} \d_\a{}^\b + {1\over 4} C_{ab}{}^{\gb}
(\g^{ab})_\a{}^\b,\quad \Ct_{\ab}{}^{\bb\g} = \Ct^\g
\d_{\ab}{}^{\bb} + {1\over 4} \Ct_{ab}{}^\g
(\g^{ab})_{\ab}{}^{\bb},$$
$$
S_{\a\ab}{}^{\b\bb} = S \d_\a{}^\b \d_{\ab}{}^{\bb} + {1\over 4}
S_{ab} (\g^{ab})_\a{}^\b \d_{\ab}{}^{\bb} + {1\over 4} \St_{ab}
(\g^{ab})_{\ab}{}^{\bb} \d_\a{}^\b + {1\over{16}} S_{abcd}
(\g^{ab})_\a{}^\b (\g^{cd})_{\ab}{}^{\bb}.$$

The action of \action\ is BRST invariant if the background fields
satisfy suitable constraints. As was shown in \BerkovitsUE, these
constraints imply that the background field satisfy the type II
supergravity equations. The BRST invariance is obtained by requiring
that the BRST currents $j_B = \l^\a d_\a$ and $\jt_B = \lt^{\ab}
\dt_{\ab}$ are conserved. Besides, the BRST charges $Q = \oint j_B$
and $\Qt = \oint \jt_B$ are nilpotent and anticommute. Let us review
these properties now.

\subsec{Nilpotency}

As was shown in \BerkovitsUE\ (see also
\ref\ChandiaIX{O.~Chand\'{\i}a, ``A note on the Classical BRST
Symmetry of the Pure Spinor String in a Curved Background,'' JHEP
0607 (2006) 019 [arXiv:hep-th/0604115].}), nilpotency is obtained
after defining momentum variables in \action\ and then using the
canonical Poisson brackets. The only momentum variable that does
not appear in \action\ is the conjugate momentum of $Z^M$ which is
defined as $P_M = (2\pi\a') \d S / \d (\p_0 Z^ M)$ where $\p_0 =
\ha ( \p + \pb )$. It is not difficult to see that $\o_\a$ is the
conjugate momentum to $\l^\a$ and that $\ot_{\ab}$ is the one for
$\lt^{\ab}$. Nilpotence of $Q$ determines the constraints

\eqn\qdos{\l^\a \l^\b H_{\a\b A} = \l^\a \l^\b \l^\g R_{\a\b\g}{}^\d
 = \l^\a \l^\b \Rt_{\a\b\gb}{}^{\db} = 0,}
$$
\l^\a \l^\b T_{\a\b}{}^a = \l^\a \l^\b T_{\a\b}{}^\g = \l^\a \l^\b
T_{\a\b}{}^{\gb} = 0,$$ where $H=dB$, the torsion $T_{AB}{}^\a$
and $R_{AB\g}{}^\d$ are the torsion and the curvature constructed
using $\O_{A\b}{}^\g$ as connection. Similarly, $T_{AB}{}^{\gb}$
and $\Rt_{AB\gb}{}^{\db}$ are the torsion and the curvature using
$\Ot_{A\bb}{}^{\gb}$ as connection.

The nilpotence of the BRST charge $\Qt$ leads to the constraints

\eqn\qtildos{\lt^{\ab} \lt^{\bb} H_{\ab\bb A} = \lt^{\ab}
\lt^{\bb} R_{\ab\bb\g}{}^\d = \lt^{\ab} \lt^{\bb} \lt^{\gb}
\Rt_{\ab\bb\gb}{}^{\db} = 0,}
$$
\lt^{\ab} \lt^{\bb} T_{\ab\bb}{}^a = \lt^{\ab} \lt^{\bb}
T_{\ab\bb}{}^\g = \lt^{\ab} \lt^{\bb} \Tt_{\ab\bb}{}^{\gb} = 0.$$
Finally, the anticommutation between $Q$ and $\Qt$ determines

\eqn\qqtil{H_{\a\bb A} = T_{\a\bb}{}^a = T_{\a\bb}{}^\g =
T_{\a\bb}{}^{\gb} = \l^\a \l^\b R_{\gb\a\b}{}^\d = \lt^{\ab}
\lt^{\bb} \Rt_{\g\ab\bb}{}^{\db} = 0.}

Note that given the decomposition \fields\ for the connections, we
can respectively write

\eqn\decurvatures{R_{DC\a}{}^\b = R_{DC}\d_\a{}^\b + {1\over 4}
R_{DCef} (\g^{ef})_\a{}^\b,} $$ \Rt_{DC\ab}{}^{\bb} =
\Rt_{DC}\d_{\ab}{}^{\bb} + {1\over
4}\Rt_{DCef}(\g^{ef})_{\ab}{}^{\bb}.$$

\subsec{Holomorphicity}

The holomorphicity of $j_B$ and the antiholomorphicity of $\jt_B$
constraints are determined after the use of the equations of motion
derived from the action \action. The equation for the pure spinor
ghosts are

\eqn\pureseom{\Nb\l^\a + \l^\b ( \dt_{\gb} C_\b{}^{\a\gb} +
\lt^{\ab} \ot_{\bb} S_{\b\ab}{}^{\a\bb} ) = 0,\quad \Nb\o_\a - (
\dt_{\gb} C_\a{}^{\b\gb} + \lt^{\ab} \ot_{\bb} S_{\a\ab}{}^{\b\bb}
)\o_\b = 0, } and

\eqn\gheom{\N\lt^{\ab} + \lt^{\bb} ( d_\g \Ct_{\bb}{}^{\ab\g} +
\l^\a \o_\b S_{\a\bb}{}^{\b\ab} ) = 0,\quad \N\ot_{\ab }- ( d_\g
\Ct_{\ab}{}^{\bb\g} + \l^\a \o_\b S_{\a\ab}{}^{\b\bb} )\ot_{\bb} =
0, } where $\N$ is a covariant derivative which acts with $\O$ or
$\Ot$ connections according to the index structure of the fields it
is acting on. For example,

$$
\N P^{\a\bb} = \p P^{\a\bb} + P^{\g\bb} \O_\g{}^\a + P^{\a\gb}
\Ot_{\gb}{}^{\bb}.$$ The variations respect to $d_\a$ and
$\dt_{\ab}$ provide the equations

\eqn\pieom{\Pb^\a + \dt_{\bb} P^{\a\bb} + \lt^{\ab} \ot_{\bb}
\Ct_{\ab}{}^{\bb\a} = 0,\quad \Pi^{\ab} - d_\b P^{\b\ab} + \l^\a
\o_\b C_\a{}^{\b\ab} = 0.} The most difficult equations to obtain
are those coming from the variation of the superspace coordinates.
Let us define $\s^A = \d Z^M E_M{}^A$, then it is not difficult to
obtain

$$
\d \Pi^A = \p \s^A - \s^B \Pi^C E_B{}^M E_C{}^N \p_{[N} E_{M]}{}^A
(-1)^{C(B+M)}.$$ Here we can express this variation in terms of
the connection $\O$ . In fact,

$$
\d \Pi^A = \N \s^A - \s^B \Pi^C ( T_{CB}{}^A + \O_{BC}{}^A
(-1)^{BC} ). $$
There is a point about our notation for the torsion that we should make
clear. Using tangent superspace indices, the torsion can be written
as
\eqn\TorsionI{T_{BC}{}^A = -E_B{}^N (\p_N E_C{}^M)E_M{}^A +(-)^{BC}E_C{}^N
(\p_N E_B{}^M)E_M{}^A +\O_{BC}{}^A -(-)^{BC}\O_{CB}{}^A .}
In our notation, $T_{BC}{}^\a$ will mean that the connection in \TorsionI\
is $\O_{C\b}{}^\a$ while $T_{BC}{}^{\ab}$ means that the connection in \TorsionI\
 is $\Ot_{C\bb}{}^{\ab}$. Since we also have two connections with bosonic
tangent space index $\O_{Cb}{}^a$ and $\Ot_{Cb}{}^a$, we use $T_{BC}{}^a$ to
denote the torsion when we use the first and $\Tt_{BC}{}^a$ to denote the
torsion when we use the second.

We vary the action \action\ under these transformations and, after
using the equations \gheom, \pieom\ and some of the nilpotence
constraints, we obtain

\eqn\deom{\Nb d_\a = -\ha \Pi^a \Pb^b ( T_{\a(ba)} + H_{\a ba} ) +
\ha \Pi^\b \Pb^a ( T_{\b\a a} - H_{\b\a a} ) - d_\b \Pb^a
T_{a\a}{}^\b} $$ - \dt_{\bb} \Pi^a ( T_{a\a}{}^{\bb} + \ha
P^{\g\bb} ( T_{\g\a a} + H_{\g\a a} ) ) + \l^\b \o_\g \Pb^a
R_{a\a\b}{}^\g + \lt^{\bb} \ot_{\gb} \Pi^a ( \Rt_{a\a\bb}{}^{\gb} -
\ha \Ct_{\bb}{}^{\gb\d} ( T_{\d\a a} + H_{\d\a a} ) )$$ $$ -
\dt_{\bb} \Pi^\g ( T_{\g\a}{}^{\bb} + \ha P^{\d\bb} H_{\d\g\a} ) +
\l^\b \o_\g \Pb^{\db} R_{\db\a\b}{}^\g + \lt^{\bb} \ot_{\gb} \Pi^\d
( \Rt_{\d\a\bb}{}^{\gb} + \ha \Ct_{\bb}{}^{\gb\r} H_{\r\d\a} )$$
$$
+ d_\b \dt_{\gb} ( P^{\d\gb} T_{\d\a}{}^\b - \N_\a P^{\b\gb} ) +
\lt^{\bb} \ot_{\gb} d_\d ( \N_\a \Ct_{\bb}{}^{\gb\d} +
\Ct_{\bb}{}^{\gb\r} T_{\r\a}{}^\d + P^{\d\rb}
\Rt_{\rb\a\bb}{}^{\gb} )$$
$$
+\l^\b \o_\g \dt_{\db} ( \N_\a C_\b{}^{\g\db} - P^{\r\db}
R_{\r\a\b}{}^\g ) - \l^\b \o_\g \lt^{\db} \ot_{\rb} ( \N_\a
S_{\b\db}{}^{\g\rb} + C_\b{}^{\g\sb} \Rt_{\sb\a\db}{}^{\rb} +
\Ct_{\db}{}^{\rb\s} R_{\s\a\b}{}^\g ),$$ and

\eqn\dteom{\N \dt_{\ab} = -\ha \Pi^a \Pb^b ( T_{\ab(ba)} + H_{\ab
ba} ) + \ha \Pi^a \Pb^{\bb} ( T_{\bb\ab a} + H_{\bb\ab a} ) -
\dt_{\bb} \Pi^a T_{a\ab}{}^{\bb} } $$ - d_\b \Pb^a (
T_{a\ab}{}^\b - \ha P^{\b\gb} ( T_{\gb\ab a} - H_{\gb\ab a} ) ) +
\lt^{\bb} \ot_{\gb} \Pi^a \Rt_{a\ab\bb}{}^{\gb} + \l^\b \o_\g
\Pb^a ( R_{a\ab\b}{}^\g - \ha C_\b{}^{\g\db} ( T_{\db\ab a} -
H_{\db\ab a} ) )$$ $$ - d_\b \Pb^{\gb} ( T_{\gb\ab}{}^\b + \ha
P^{\b\db} H_{\gb\db\ab} ) + \lt^{\bb} \ot_{\gb} \Pi^\d
\Rt_{\d\ab\bb}{}^{\gb} + \l^\b \o_\g \Pb^{\db} ( R_{\db\ab\b}{}^\g
+ \ha C_\b{}^{\g\rb} H_{\db\rb\ab} )$$
$$
+ d_\b \dt_{\gb} ( P^{\b\db} T_{\db\ab}{}^{\gb} - \N_{\ab}
P^{\b\gb} ) + \l^\b \o_\g \dt_{\db} ( \N_{\ab} C_\b{}^{\g\db} +
C_\b{}^{\g\rb} \Tt_{\rb\ab}{}^{\db} - P^{\r\db} R_{\r\ab\b}{}^\g
)$$
$$
+\lt^{\bb} \ot_{\gb} d_\d ( \N_{\ab} \Ct_{\bb}{}^{\gb\d} + P^{\d\rb}
\Rt_{\rb\ab\bb}{}^{\gb} ) - \l^\b \o_\g \lt^{\db} \ot_{\rb} (
\N_{\ab} S_{\b\db}{}^{\g\rb} + C_\b{}^{\g\sb}
\Rt_{\sb\ab\db}{}^{\rb} + \Ct_{\db}{}^{\rb\s} R_{\s\ab\b}{}^\g ).$$
From these equations, \pureseom, \gheom\ and also two equations
in \qqtil\ we obtain the holomorphicity constraints. In fact, $\Nb
j_B = 0$ implies

\eqn\pjb{ T_{\a(ab)} = H_{\a ab} = T_{\a\b a} - H_{\a\b a} =
T_{a\a}{}^\b = T_{a\a}{}^{\bb} + P^{\g\bb} T_{\g\a a} = \l^\a
\l^\b R_{a\a\b}{}^\g = 0,}
$$
\Rt_{a\a\bb}{}^{\gb} - \Ct_{\bb}{}^{\gb\d} T_{\d\a a} =
T_{\g\a}{}^{\bb} + \ha P^{\d\bb} H_{\d\g\a} =
\Rt_{\d\a\bb}{}^{\gb} + \ha \Ct_{\bb}{}^{\gb\r} H_{\r\d\a} =
P^{\d\gb} T_{\d\a}{}^\b - \N_\a P^{\b\gb} - C_\a{}^{\b\gb} =  0,
$$
$$
\N_\a \Ct_{\bb}{}^{\gb\d} + \Ct_{\bb}{}^{\gb\r} T_{\r\a}{}^\d +
P^{\d\rb} \Rt_{\rb\a\bb}{}^{\gb} - S_{\a\bb}{}^{\d\gb} = \l^\a \l^\b
( \N_\a C_\b{}^{\g\db} - P^{\r\db} R_{\r\a\b}{}^\g ) = 0,$$ $$ \l^\a
\l^\b ( \N_\a S_{\b\db}{}^{\g\rb} + C_\b{}^{\g\sb}
\Rt_{\sb\a\db}{}^{\rb} + \Ct_{\db}{}^{\rb\s} R_{\s\a\b}{}^\g ) =
0,$$ and $\N \jt_B = 0$ implies

\eqn\pjtb{T_{\ab(ab)} = H_{\ab ab} = T_{\ab\bb a} + H_{\ab\bb a} =
T_{a\ab}{}^{\bb} = T_{a\ab}{}^\b - P^{\b\gb} T_{\gb\ab a} =
\lt^{\ab} \lt^{\bb} \Rt_{a\ab\bb}{}^{\gb} = 0,}
$$
R_{a\ab\b}{}^\g - C_\b{}^{\g\db} T_{\db\ab a} = T_{\gb\ab}{}^\b +
\ha P^{\b\db} H_{\gb\db\ab} = R_{\db\ab\b}{}^\g + \ha C_\b{}^{\g\rb}
H_{\db\rb\ab} = P^{\b\db} T_{\db\ab}{}^{\gb} - \N_{\ab} P^{\b\gb}
+ \Ct_{\ab}{}^{\gb\b} =  0,
$$
$$
\N_{\ab} C_\b{}^{\g\db} + C_\b{}^{\g\rb} T_{\rb\ab}{}^{\db} -
P^{\r\db} R_{\r\ab\b}{}^\g - S_{\b\ab}{}^{\g\db} = \lt^{\ab}
\lt^{\bb} ( \N_{\ab} \Ct_{\bb}{}^{\gb\d} + P^{\d\rb}
\Rt_{\rb\ab\bb}{}^{\gb} ) = 0,$$
$$ \lt^{\ab} \lt^{\bb} ( \N_{\ab} S_{\d\bb}{}^{\r\gb} + C_\d{}^{\r\sb}
\Rt_{\sb\ab\bb}{}^{\gb} + \Ct_{\bb}{}^{\gb\s} R_{\s\ab\d}{}^\r ) =
0.$$

\subsec{Solving the Bianchi identities}

We can gauge-fix some of the torsion components and determine others
through the use of Bianchi identities. It is not necessary but it
will simplify the computation of the one-loop beta functions. As in
\BerkovitsUE, we can set $H_{\a\b\g} = H_{\a\b\gb} = H_{\a\bb\gb} =
H_{\ab\bb\gb} = 0$ since there is no such ten-dimensional
superfields satisfying the constrains of \pjb\ and \pjtb. We can use
the Lorentz rotations to gauge fix $T_{\a\b}{}^a = \g^a_{\a\b}$ and
$T_{\ab\bb}{}^a = \g^a_{\ab\bb}$, therefore the above constraints
imply $H_{\a\b a} = (\g_a)_{\a\b}$ and $H_{\ab\bb a} =
-(\g_a)_{\ab\bb}$. We can use the shift symmetry of the action
\action\

$$
\d d_\a = \d \O_{\a\b}{}^\g \l^\b \o_\g, \quad \d \dt_{\ab} = \d
\Ot_{\ab\bb}{}^{\gb} \lt^{\bb} \ot_{\gb},\quad \d C_\a{}^{\b\gb} =
P^{\d\gb} \d \O_{\d\a}{}^\b,\quad \d \Ct_{\ab}{}^{\bb\g} = -
P^{\g\db} \d \Ot_{\db\ab}{}^{\bb},
$$
$$
\d S_{\a\bb}{}^{\g\db} = C_\a{}^{\g\rb} \d \Ot_{\rb\bb}{}^{\db} +
\Ct_{\bb}{}^{\db\r} \d \O_{\r\a}{}^\g,$$ to gauge-fix $T_{\a\b}{}^\g
= T_{\ab\bb}{}^{\gb} = 0$.

The Bianchi identity for the torsion is

\eqn\bianchis{(\N T)_{ABC}{}^D \equiv \N_{[A} T_{BC]}{}^D +
T_{[AB}{}^E T_{EC]}{}^D - R_{[ABC]}{}^D = 0,}
 where brackets in \bianchis\ mean (anti-)symmetrization
respect to the $ABC$ indices. The curvature will be $R$ or $\Rt$
if the upper index $D$ is $\d$ or $\db$ respectively. When
$D = d$, we use the notation $(\N T)_{ABC}{}^d$ or $(\N \Tt)_{ABC}{}^d$,
if we use the connection $\O_{Bc}{}^a$ or $\Ot_{Bc}{}^a$;
then the curvatures in each case will be $R$ or $\Rt$.

The Bianchi identity $(\N T)_{\a\b\g}{}^a = 0$ implies $T_{\a ab} =
2 (\g_{ab})_\a{}^\b \O_\b$. Similarly, the Bianchi identity $(\N
\Tt)_{\ab\bb\gb}{}^a = 0$ implies $\Tt_{\ab ab} = 2
(\g_{ab})_{\ab}{}^{\bb} \Ot_{\bb}$. The Bianchi identity $(\N
T)_{\a\bb\gb}{}^a = 0$ implies $\Ot_\a = \Tt_{\a a}{}^b = 0$.
Similarly, the Bianchi identity $(\N T)_{\ab\b\g}{}^a = 0$ implies
$\O_{\ab} = T_{\ab a}{}^b = 0$. It is not difficult to show that the
constraints $T_{a\a}{}^\a = T_{a\ab}{}^{\ab} = 0$ imply $\O_a =
\Ot_a = 0$.

We can write two sets of Bianchi identities for $H$ depending on
what is the connection we choose in the covariant derivative. Note
that the components of the superfield $H$ do not depend on such
choice. The Bianchi identities come from
$\N H = 0$ and $\Nt H = 0$ and it is not difficult to check that
both sets are equivalent. Let us write only one of
them

\eqn\dhzero{(\N H)_{ABCD} \equiv \N_{[A} H_{BCD]} + {3\over 2}
T_{[AB}{}^E H_{ECD]} = 0.} There is one more Bianchi identity
involving a derivative of the curvature \eqn\drs{(\N R)_{ABCD}{}^E
\equiv \N R_{[ABC]D}{}^E + T_{[AB}{}^F R_{F C]D}{}^E=0.}

The identities $(\N H)_{\a\b\g\d}, (\N
H)_{\a\b\g\db}, (\N H)_{\a\b\gb\db}, (\N H)_{\a\bb\gb\db}, (\N
H)_{\ab\bb\gb\db}$ are easily satisfied if we recall the identities
for gamma matrices $\g^a_{(\a\b} (\g_a)_{\g)\d} = \g^a_{(\ab\bb}
(\g_a)_{\gb)\db} = 0$. The identities $(\N H)_{a\a\b\g}, (\N
H)_{a\a\b\gb}, (\N H)_{a\a\bb\gb}, (\N H)_{a\ab\b\gb}$ are satisfied
after using the dimension-$\ha$ constraints. The identity $(\N
H)_{ab\a\b} = 0$ implies $T_{abc} + H_{abc} = 0$ and the identity
$(\Nt H)_{ab\ab\bb} = 0$ implies $\Tt_{abc} - H_{abc} = 0$. The
identity $(\N H)_{ab\a\bb} = 0$ is satisfied if we use the
constraints involving the superfield $P^{\a\bb}$ in the first lines
of \pjb\ and \pjtb.

\subsec{The remaining equation of motion}

In the computation of the one-loop beta function we will need to
know the equation of motion for $\Pi^a$ and $\Pb^a$. Since we know
that the difference $\N \Pb^a - \Nb \Pi^a$ is given by the torsion
components, then we only need to determine $\N \Pb^a + \Nb \Pi^a$
which is determined by the varying the action respect to $\s^a = \d
Z^M E_M{}^a$. To make life simpler we will write this equation using
the above results for torsion and $H$ components. The equation turns
out to be

\eqn\piaeom{\ha ( \Nt \Pb_a + \Nb \Pi_a ) = \ha \Pi^b \Pb^c H_{cba} -
\ha \Pi^\a \Pb^b T_{\a ab} + d_\a \Pb^b T_{ab}{}^\a  + \l^\a \o_\b
\Pb^b R_{ab\a}{}^\b}
$$
 + \dt_{\ab} \Pi^b ( T_{ab}{}^{\ab} + \ha
P^{\b\ab} T_{\b ab} ) + \lt^{\ab} \ot_{\bb} \Pi^b (
\Rt_{ab\ab}{}^{\bb} + \ha \Ct_{\ab}{}^{\bb\g} T_{\g ab} ) + \ha
\dt_{\ab} \Pi^\b T_{a\b}{}^{\ab} + \ha \lt^{\ab} \ot_{\bb} \Pi^\g
\Rt_{a\g\ab}{}^{\bb}$$ $$ + \ha d_\a \Pb^{\bb} T_{a\bb}{}^\a + \ha
\l^\a \o_\b \Pb^{\gb} R_{a\gb\a}{}^\b + d_\a \dt_{\bb} \N_a
P^{\a\bb} + \l^\a \o_\b \dt_{\gb} ( \N_a C_\a{}^{\b\gb} - P^{\d\gb}
R_{a\d\a}{}^\b )$$ $$ + \lt^{\ab} \ot_{\bb} d_\g ( \N_a
\Ct_{\ab}{}^{\bb\g} + P^{\g\db} \Rt_{a\db \ab}{}^{\bb} ) + \l^\a
\o_\b \lt^{\gb} \ot_{\db} ( \N_a S_{\a\gb}{}^{\b\db} -
\Ct_{\gb}{}^{\db\r} R_{a\r\a}{}^\b - C_\a{}^{\b\rb}
\Rt_{a\rb\gb}{}^{\db} ).$$

\subsec{Ghost number conservation}

As it was shown in \BerkovitsUE, the vanishing of the ghost number
anomaly determines that the spinorial derivatives of the dilaton
superfield $\Phi$ are proportional to the connection $\O$. This
relation is crucial to cancel the beta function in heterotic string
case \ChandiaHN\ and will be equally essential in our computation.
Let us recall how this relation is obtained. Consider the coupling
between ghost number currents and the connections in the action
\action. Namely

$$
{1\over{2\pi\a'}} \int d^2z ~ (J \Ob + \Jt \O ).$$ The BRST variation
on this term contains the term

$$
- {1\over{2\pi\a'}} \int d^2z ~ (\pb J \l^\a \O_\a + \p \Jt \l^\a
\Ot_\a ).$$ The anomaly in the ghost number current conservation turns
out to be proportional to the two dimensional Ricci scalar, as noted
by dimensional grounds. The proportionality can be determined by
performing a Weyl transformation, around the flat world-sheet, of
the anomaly equation. In this way, the triple-pole in the OPE
between the current and the corresponding stress tensor yields

\eqn\dilat{\N_\a \Phi = 4 \O_\a,\quad \N_{\ab} \Phi = 4 \Ot_{\ab},}
which will be used in section 5 to cancel the UV divergent part of
the effective action.

\newsec{Covariant Background Field Expansion}

We use the method explained in \ref\Skenderis{J. de Boer and K.
Skenderis, ``Covariant Computation of the Low Energy Effective
Action of the Heterotic Superstring,'' Nucl. Phys. B481 (1996) 129
[arXiv:hep-th/9608078].} and \ChandiaHN. Here, we need to define a
straight-line geodesic which joins a point in superspace to
neighbor ones and allows us to perform an expansion in superspace.
It is given by $Y^A$ which satisfies the geodesic equation $\D Y^A =
Y^B \N_B Y^A = 0$. The connection we choose to define this covariant
derivative has the non-vanishing components $\O_{Aa}{}^b,
\O_{A\a}{}^\b$ and $\Ot_{A\ab}{}^{\bb}$. These same connections
are defined in the action \action. In this way, the covariant
expansions of the different objects in \action\ are determined by

\eqn\covexps{\D \Pi^A = \N Y^A - Y^B \Pi^C T_{CB}{}^A,\quad \D
\Ob_\a{}^\b = -Y^A \Pb^B R_{BA\a}{}^\b, \quad \D \Ot_{\ab}{}^{\bb} =
-Y^A \Pi^B \Rt_{BA\ab}{}^{\bb}.} Any superfield $\Psi$ is expanded
as $\D \Psi = Y^A \N_A \Psi$.

As in \ChandiaHN, we see that $d_{\a}, \dt_{\ab}$ and the pure
spinor ghosts are treated as fundamental fields, then we expand them
according to

\eqn\fund{ d_{\a} = d_{\a 0} + \dh_{\a},\quad \l^{\a} = \l_0^{\a} +
\lh^{\a}, \quad \o_{\a} = \o_{\a 0} + \oh_{\a},}
$$
\dt_{\ab} = \dt_{\ab0} + \dth_{\ab},\quad \lt^{\ab} = \lt_0^{\ab} +
\lth^{\ab},\quad \ot_{\ab} = \ot_{\ab 0} + \oth_{\ab 0},$$ where
the subindex $0$ means the background value of the corresponding
field which will dropped in the subsequent discussion.

The quadratic part of the expansion of \action, excluding the
Fradkin-Tseytlin term, has the form

\eqn\sdos{ S_2 = S_p + {1\over{2\pi\a'}} \int d^2z ~ (Y^A Y^B E_{BA}
+ Y^A \Nb Y^B C_{BA} + Y^A \N Y^B \Cb_{BA} + \dh_\a Y^A \Db_A{}^\a}
$$ + \dth_{\ab} Y^A D_A{}^{\ab}   + (\lh^\a \oh_\b) \Hb_\a{}^\b +
(\lth^{\ab} \oth_{\bb}) H_{\ab}{}^{\bb} + (\lh^\a \o_\b + \l^\a
\oh_\b) Y^A \Ib_{A\a}{}^\b + (\lth^{\ab} \ot_{\bb} + \lt^{\ab}
\oth_{\bb}) Y^A I_{A\ab}{}^{\bb} $$ $$ + \dh_\a \dth_{\bb} P^{\a\bb}
 + (\lh^\a \o_\b + \l^\a \oh_\b) \dth_{\gb} C_\a{}^{\b\gb} +
(\lth^{\ab} \ot_{\bb} + \lt^{\ab} \oth_{\bb}) \dh_\g
\Ct_{\ab}{}^{\bb\g} + (\lh^\a \o_\b + \l^\a \oh_\b) (\lth^{\gb}
\ot_{\db} + \lt^{\gb} \oth_{\db}) S_{\a\gb}{}^{\b\db} ) ,$$ where
$E_{BA}, C_{BA}, \dots$ are background superfields given by

\eqn\eba{E_{BA} = {1\over 4} \Pi^C \Pb^D ( T_{CB}{}^E H_{EDA}
(-1)^{D(C+B)} - T_{DB}{}^E H_{ECA} (-1)^{BC} } $$ + \N_B
H_{DCA}(-1)^{B(C+D)}  + 2 T_{CB}{}^a T_{DAa} (-1)^{D(C+B)} )-
{1\over 4} \Pi^{(a} \Pb^{C)} ( R_{CBAa} - T_{CB}{}^D T_{DAa}$$ $$ +
\N_B T_{CAa} (-1)^{BC} ) + \ha d_\a \Pb^C (-1)^{A+B} ( -R_{CBA}{}^\a
+ T_{CB}{}^D T_{DA}{}^\a - \N_B T_{CA}{}^\a (-1)^{BC} )$$ $$ + \ha
\dt_{\ab} \Pi^C (-1)^{A+B} ( -R_{CBA}{}^{\ab} + T_{CB}{}^D
T_{DA}{}^{\ab} - \N_B T_{CA}{}^{\ab} (-1)^{BC} ) + \ha \l^\a \o_\b
\Pb^C ( T_{CB}{}^D R_{DA\a}{}^\b$$ $$ - \N_B R_{CA\a}{}^\b (-1)^{BC}
) + \ha \lt^{\ab} \ot_{\bb} \Pi^C ( T_{CB}{}^D R_{DA\ab}{}^{\bb} -
\N_B R_{CA\ab}{}^{\bb} (-1)^{BC} ) + \ha d_\a \dt_{\bb} \N_B \N_A
P^{\a\bb}$$ $$ + \ha \l^\a \o_\b \dt_{\gb} \N_B \N_A C_\a{}^{\b\gb}
(-1)^{A+B} + \ha \lt^{\ab} \ot_{\bb} d_\g \N_B \N_A
\Ct_{\ab}{}^{\bb\g} (-1)^{A+B} + \ha \l^\a \o_\b \lt^{\gb} \ot_{\db}
\N_B \N_A S_{\a\gb}{}^{\b\db},$$

\eqn\cba{ C_{BA} = -{1\over 4} \Pi^a T_{BAa} - \ha \Pi^C T_{CAa}
\d^a_B - {1\over 4} \Pi^A H_{CBA} - \ha d_\a T_{BA}{}^\a (-1)^{A+B}
- \ha \l^\a \o_\b R_{BA\a}{}^\b, }

\eqn\ccba{ \Cb_{BA} = -{1\over 4} \Pi^b T_{BAa} - \ha \Pb^C T_{CAa}
\d^a_B + {1\over 4} \Pi^A H_{CBA} - \ha \dt_{\ab} \Tt_{BA}{}^{\ab}
(-1)^{A+B} - \ha \lt^{\ab} \ot_{\bb} \Rt_{BA\ab}{}^{\bb},}

\eqn\dbaa{\Db_A{}^\a = - \Pb^B T_{BA}{}^\a + \dt_{\bb} \N_A
P^{\a\bb} (-1)^A + \lt^{\bb} \ot_{\gb} \N_A \Ct_{\bb}{}^{\gb\a},}

\eqn\daab{D_A{}^{\ab} = -\Pi^B \Tt_{BA}{}^{\ab} - d_\b \N_A
P^{\b\ab} (-1)^A + \l^\b \o_\g \N_A C_\b{}^{\g\ab},}

\eqn\hbab{\Hb_\a{}^\b = \Ob_\a{}^\b + \dt_{\gb} C_\a{}^{\b\gb}
\lt^{\gb} \ot_{\db} S_{\a\gb}{}^{\b\db},}

\eqn\habbb{H_{\ab}{}^{\bb} = \Ot_{\ab}{}^{\bb} + d_\g
\Ct_{\ab}{}^{\bb\g} + \l^\g \o_\d S_{\g\ab}{}^{\d\bb},}

\eqn\ibaab{\Ib_{A\a}{}^\b = -\Pb^A R_{BA\a}{}^\b + \dt_{\gb} \N_A
C_\a{}^{\b\gb} (-1)^A + \lt^{\gb} \ot_{\db} \N_A
S_{\a\gb}{}^{\b\gb},}

\eqn\iaabbb{I_{A\ab}{}^{\bb} = -\Pi^B \Rt_{BA\ab}{}^{\bb} + d_\g
\N_A \Ct_{\ab}{}^{\bb\g} (-1)^A + \l^\g \o_\d \N_A
S_{\g\ab}{}^{\d\bb}.} In \sdos\ $S_p$ provides the propagators for
the quantum fields and is given by

\eqn\sp{ S_p = {1\over{2\pi\a'}} \int d^2z ~ (\ha \N Y^a \Nb Y_a +
\dh_\a \Nb Y^\a + \dth_{\ab} \N Y^{\ab}) + {\cal L}_{pure},} where
${\cal L}_{pure}$ is the Lagrangian for the pure spinor ghosts.

\newsec{The one-loop UV divergent Part of the Effective Action}

The effective action is given by

\eqn\seff{ e^{-S_{eff}} = \int D{\cal Q} ~ e^{-S},} where ${\cal Q}$
represents the quantum fluctuations.

To compute the one-loop beta functions we need to expand \action\ up
to second order in the quantum fields. In this way, we will obtain
the UV divergent part of the effective action, $S_\L$. Here $\L$ is
UV scale. Note that the Fradkin-Tseytlin term is evaluated on a sphere
with metric $\L dz d{\bar z}$. Finally, the complete UV divergent
part of the effective action becomes

\eqn\sdiv{ S_\L + {1\over 2\pi} \int d^2 z ~ ( \N \Pb^A \N_A \Phi +
\Pb^A \Pi^B \N_B \N_A \Phi ) \log\L .}

The computation of $S_\L$ is performed by contracting the quantum
fields. From \sp\ we read

\eqn\props{ Y^a (z,{\bar z}) Y^b (w,{\bar w}) \to -\a' \eta^{ab}
\log |z-w|^2,\quad \dh_\a(z) Y^\b(w) \to
{\a'\d_\a{}^\b\over{(z-w)}},\quad \dth_{\ab}({\bar z}) Y^{\bb}({\bar
w}) \to {\a'\d_{\ab}^{\bb}\over{({\bar z}-{\bar w})}}.} For the pure
spinor ghosts we note that, because of \fields, they enter in the
combinations

$$
N^{ab} = \ha (\l \g^{ab} \o),\quad J = \l^\a \o_\a,\quad \Nn^{ab} =
\ha (\lt \g^{ab} \ot), \quad \Jt = \lt^{\ab} \ot_{\ab}.$$ We can
expand each of these combinations as $J + J_1 + J_2$, similarly for $\Jt$,
$N^{ab}$ and $\Nn^{ab}$. As in
\ChandiaHN, the only relevant OPE's involving the pure spinor ghosts
and contributing to $S_\L$ are

\eqn\nope{ N_1^{ab}(z) N_1^{cd}(w) \to {1\over(z-w)} ( - \eta^{a[c}
N^{d]b}(w) + \eta^{b[c} N^{d]a}(w) ), }

\eqn\nnope{ \Nn_1^{ab}({\bar z}) \Nn_1^{cd}({\bar w}) \to {1\over({\bar
z}-{\bar w})} ( - \eta^{a[c} \Nn^{d]b}({\bar w}) + \eta^{b[c}
\Nn^{d]a}({\bar w}) ). }

The one-loop contributions to $S_\L$ come from self-contraction of
$Y^A$'s in the term with $E_{BA}$ in \sdos\ and a series of double
contractions in \sdos. These come from products between the term
involving $C_{BA}$ with the one involving $\Cb_{BA}$, $C_{BA}$ with
$\Db_{A}{}^{\bb}$, $\Cb_{BA}$ with $D_{A}{}^{\b}$, $\Db_{A}{}^{\bb}$
with $D_{A}{}^{\b}$, $E_{BA}$ with $P^{\a\bb}$, $\Ib_{C\a}{}^{\b}$
with $C_{\a}{}^{\b\gb}$, $I_{C\ab}{}^{\bb}$ with
$\Ct_{\ab}{}^{\bb\g}$ and $S_{\a\gb}{}^{\b\db}$ with itself. After
adding up all these contributions, the one-loop UV divergent part of
the effective action is proportional to

\eqn\uv{ \int d^2z ~   [ - \eta^{ab} E_{ba} + \eta^{a[c} \eta^{d]b}
C_{ba} \Cb_{dc} + \eta^{ab} C_{[a\a]} \Db_b{}^\a + \eta^{ab}
\Cb_{[a\ab]} D_b{}^{\ab} + \Db_{\ab}{}^\b D_\b{}^{\ab} + E_{[\a\bb]}
P^{\a\bb} } $$
 + N^{ab} \Ib_{\ab a}{}^c C_{cb}{}^{\ab} + \Nn^{ab}
I_{\a a}{}^c \Ct_{cb}{}^\a + \ha N^{ab} \Nn^{cd} S_a{}^e{}_c{}^f
S_{bedf} + \N \Pb^A \N_A \Phi + \Pb^A \Pi^B \N_B \N_A \Phi ]
\log\L,$$ where we used the expressions \fields.

Now it will be shown that \uv\ vanishes as consequence of the
classical BRST constraints.

\newsec{One-loop Conformal Invariance}

To write the equations derived from the vanishing of \uv, we need to
determine $\N\Pb^A$ from the classical equations of motion from
\action. In order to do this, we need to know

\eqn\diff{\Nb \Pi^A - \N \Pb^A = \Pi^B \Pb^C T_{CB}{}^A.} Note that
we are using here the connection $\O_A{}^B$ to calculate the
covariant derivatives and the torsion components.

The equation for $\N\Pb_a$ is

\eqn\npba{\N \Pb_a = \Pi^b \Pb^c T_{abc} - \Pi^\a \Pb^b T_{\a ab} +
\dt_{\ab} \Pi^b T_{ab}{}^{\ab} + d_\a \Pb^b T_{ab}{}^\a +
\lt^{\ab} \ot_{\bb} \Pi^b \Rt_{ab\ab}{}^{\bb} + \l^\a \o_\b
\Pb^b R_{ab\a}{}^\b} $$
 + \dt_{\ab} \Pi^\b T_{a\b}{}^{\ab} + \lt^{\ab}
\ot_{\bb} \Pi^\g \Rt_{a\g\ab}{}^{\bb} + d_\a \dt_{\bb} \N_a
P^{\a\bb} + \l^\a \o_\b \dt_{\gb} ( \N_a C_\a{}^{\b\gb} - P^{\d\gb}
R_{a\d\a}{}^\b )$$ $$ + \lt^{\ab} \ot_{\bb} d_\g ( \N_a
\Ct_{\ab}{}^{\bb\g} + P^{\g\db} \Rt_{a\db \ab}{}^{\bb} ) + \l^\a
\o_\b \lt^{\gb} \ot_{\db} ( \N_a S_{\a\gb}{}^{\b\db} -
\Ct_{\gb}{}^{\db\r} R_{a\r\a}{}^\b - C_\a{}^{\b\rb}
\Rt_{a\rb\gb}{}^{\db} ).$$ Now we compute the equation for $\Pb^\a$.
We start by noting that this world-sheet field is determined from
the equation of motion \pieom, then

$$
\N\Pb^\a = - \N ( \dt_{\bb} P^{\a\bb} + \lt^{\bb} \ot_{\gb}
\Ct_{\bb}{}^{\gb\a} ).$$

Remember that the covariant derivative on $P^{\a\bb}$ and
$\Ct_{\ab}{}^{\bb\g}$ acts with $\O_\a{}^\b$ on $\a$-indices and with
$\Ot_{\ab}{}^{\bb}$ on ${\ab}$-indices. Now we can use the equations
\gheom\ and \dteom\ to obtain

\eqn\npbaa{ \N \Pb^\a = d_\b \dt_{\gb} ( \Ct_{\db}{}^{\gb\b}
P^{\a\db} + P^{\b\db} \N_{\db} P^{\a\gb} ) + \l^\b \o_\g \dt_{\db} (
- S_{\b\rb}{}^{\g\db} P^{\a\rb} + C_\b{}^{\g\rb} \N_{\rb} P^{\a\db}
)} $$ - \dt_{\bb} \Pi^a \N_a P^{\a\bb}
 - \dt_{\bb} \Pi^\g \N_\g
P^{\a\bb} + \lt^{\bb} \ot_{\gb} d_\d ( \Ct_{\bb}{}^{\rb\d}
\Ct_{\rb}{}^{\gb\a} - \Ct_{\rb}{}^{\gb\d} \Ct_{\bb}{}^{\rb\a} -
P^{\d\rb} \N_{\rb} \Ct_{\bb}{}^{\gb\a}$$ $$
-P^{\a\db}(\N_{\db}\Ct_{\bb}{}^{\gb\d}+P^{\d\eb}\Rt_{\eb\db\bb}{}^{\gb})
) + \l^\b \o_\g \lt^{\db} \ot_{\rb} ( S_{\b\db}{}^{\g\sb}
\Ct_{\sb}{}^{\rb\a} - S_{\b\sb}{}^{\g\rb} \Ct_{\db}{}^{\sb\a} +
C_\b{}^{\g\sb} \N_{\sb} \Ct_{\db}{}^{\rb\a}+P^{\a\eb}(\N_{\eb}
S_{\b\db}{}^{\g\rb}$$ $$ + C_\b{}^{\g\sb} \Rt_{\sb\eb\db}{}^{\rb} +
\Ct_{\db}{}^{\rb\s} R_{\s\eb\b}{}^\g )) - \lt^{\bb} \ot_{\gb} \Pi^a
(\N_a \Ct_{\bb}{}^{\gb\a} +\Rt_{a\db\bb}{}^{\gb}P^{\a\db}) -
\lt^{\bb} \ot_{\gb} \Pi^\d S_{\d\bb}{}^{\a\gb}.$$ To obtain the
equation for $\Pb^{\ab}$ we can use \diff. After all this we get

\eqn\npbba{ \N \Pb^{\ab} = d_\b
\dt_{\gb} ( C_\d{}^{\b\gb} P^{\d\ab} - P^{\d\gb} \N_\d P^{\b\ab} )
+ \lt^{\bb} \ot_{\gb} d_\d ( S_{\r\bb}{}^{\d\gb} P^{\r\ab} -
\Ct_{\bb}{}^{\gb\r} \N_\r P^{\d\ab} )} $$+ d_\b \Pb^a \N_a
P^{\b\ab} + d_\b \Pb^{\gb} \N_{\gb} P^{\b\ab} + \l^\b \o_\g
\dt_{\db} ( C_\b{}^{\r\db} C_\r{}^{\g\ab} - C_\r{}^{\g\db}C_\b{}^{\r\ab} +
P^{\r\db} \N_\r C_\b{}^{\g\ab}
$$ $$+P^{\d\ab}( \N_\d C_\b{}^{\g\db} - P^{\e\db} R_{\e\d\b}{}^\g )) + \l^\b \o_\g \lt^{\db}
\ot_{\rb} ( S_{\b\db}{}^{\s\rb} C_\s{}^{\g\ab} - S_{\s\db}{}^{\g\rb}
C_\b{}^{\s\ab} + \Ct_{\db}{}^{\rb\s} \N_\s C_\b{}^{\g\ab}$$ $$
+P^{\a\eb}(\N_{\eb} S_{\b\db}{}^{\g\rb} + C_\b{}^{\g\gb}
\Rt_{\gb\eb\db}{}^{\rb} + \Ct_{\db}{}^{\rb\s} R_{\s\eb\b}{}^\g)) -
\l^\b \o_\g \Pb^a (\N_a C_\b{}^{\g\ab} -R_{a\e\b}{}^{\g}P^{\e\ab})$$
$$ - \l^\b \o_\g \Pb^{\db} S_{\b\db}{}^{\g\ab} + \Pi^a \Pb^b
T_{ab}{}^{\ab} - \Pi^\b \Pb^a T_{a\b}{}^{\ab} - \dt_{\bb} \Pi^a
P^{\g\bb} T_{a\g}{}^{\ab} - \lt^{\bb} \ot_{\gb} \Pi^a
\Ct_{\bb}{}^{\gb\d} T_{a\d}{}^{\ab}.$$

\subsec{Beta functions}

Now we can obtain the equations for the background fields implied
by the vanishing of the beta functions. These are the background
dependent expressions for the conformal weights $(1,1)$
independent couplings in \uv. That is, all the independent
combinations formed from the products between $ (
\Pi^a, \Pi^\a, d_\a, \l^\a \o_\b ) $ and $ ( \Pb^a, \Pb^{\ab},
\dt_{\ab}, \lt^{\ab} \ot_{\bb} ) $ because $\Pi^{\ab}$ and
$\Pb^\a$ are determined from the equations of motion \pieom.
Let
us first concentrate on the beta functions coming from the couplings
to $\Pi^A \Pb^B$,  $d_\a \Pb^B$ and $\Pi^A \dt_{\bb}$ fields.
After using the results for the expansion \eba\--\iaabbb\ and the
equations \npba\--\npbba\ in \uv, the couplings $\Pi^\a \Pb^{\bb},
\Pi^\a \Pb^{b}, \Pi^a \Pb^{\bb}$ and $\Pi^a \Pb^b$ lead respectively
to a first set of equations

\eqn\pipidos{T_{c\bb}{}^\d T_{\d\a}{}^c-T_{c\a}{}^{\db}T_{\db
\bb}{}^c + 4\N_{\a}\N_{\bb}\F =0,}

\eqn\pipitres{\N^d T_{\a db}+R_{\a deb}\eta^{de}+T_{bc}{}^{\d}
T_{\d\a}{}^c +4\N_b \N_\a \F =0,}

\eqn\pipicuatro{R_{\bb dea}\eta^{de}+T_{ac}{}^{\db} T_{\db\bb}{}^c
-T_{c\bb}{}^\d T_{\d a}{}^c +4\N_a\N_{\bb} \F =0,}

\eqn\pipiuno{\eta^{cd} ( R_{acdb} + R_{bcda} ) - \N^c T_{abc} +
T_{c(a}{}^\a T_{b)\a}{}^c + 8 T_{a\a}{}^{\bb} T_{b\bb}{}^\a + 4
T_{ab}{}^c \N_c \Phi + 4 T_{ab}{}^{\ab}\N_{\ab}\F + 4 \N_a \N_b \Phi
= 0.} We wrote them by increasing their dimensions, that is, if
$X^a$ has dimension $-1$ and each $\theta ^\a$, $\tt^{\ab}$ have
dimension $-{1\over 2}$, then the first has dimension $1$, the
second and third dimension ${3\over 2}$ and the fourth dimension
$2$. The couplings to $d_\a \Pb^{\bb}$, $\Pi^\a \dt_{\bb}$,
$d_\a \Pb^b$ and $\Pi^a \dt_{\bb}$ lead respectively to a second set of
equations

\eqn\dalPbbb{\N^c T_{c\bb}{}^{\a} - 2 \N_{\bb} P^{\a\gb} \N_{\gb} \F
+ 2 P^{\a\gb} \N_{\gb} \N_{\bb} \F = 0,}

\eqn\Pialdtbb{\N^c T_{c\a}{}^{\bb} + 2 \N_{\a} P^{\g\bb} \N_{\g} \F
- 2 P^{\g\bb} \N_\g \N_\a \F = 0,}

\eqn\dalPbb{\N^c T_{cb}{}^{\a} - T_{cd}{}^{\a} T_{b}{}^{cd} + (T_{\d
b}{}^c T_{c\gb}{}^{\a} - R_{b\gb\d}{}^{\a}) P^{\d\gb} +
T_{b\gb}{}^{\d}( 3 \N_{\d} P^{\a\gb} - 2 P^{\a\gb}\N_{\d} \F )$$
$$ + 2 T_{bc}{}^{\a} \N^c \F - 2 \N_{b} P^{\a\gb} \N_{\gb} \F = 0,}

\eqn\Piadtbb{\N^c T_{ca}{}^{\bb} - 2 T_{cd}{}^{\bb} T_{a}{}^{cd} +
P^{\g\bb} T_{\a}{}^{de} T_{ade} + \Rt_{a\g\db}{}^{\bb} P^{\g\db} -
T_{a\g}{}^{\db} ( 3 \N_{\db} P^{\g\bb} - 2 P^{\g\bb}\N_{\db} \F )$$
$$ + 2 T_{ac}{}^{\bb} \N^c \F + 2 \N_{a} P^{\g\bb} \N_{\g} \F = 0.}
The first two with have dimension 2 and the second two have dimension ${5\over 2}$.
Now we will prove that these equations are implied by the classical
BRST constraints, the Bianchi identities \bianchis\ and the
relations \dilat.

Firstly, it is important to know the expression for the scale
curvature in terms of the scale connection. This are found to be

\eqn\curcon{ R_{\a\b} = \N_{(\a } \O_{\b )},  \quad R_{\a\bb} =
\N_{\bb} \O_\a ,\quad R_{\ab\bb} =0, $$ $$ R_{ab} = T_{ab}{}^\g
\O_\g ,\quad R_{a\b} = \N_a \O_\b ,\quad R_{a\bb} = T_{a\bb}{}^\g
\O_\g. }

\eqn\curcont{ \Rt_{\ab\bb} = \N_{(\ab }\Ot_{\bb )},  \quad
\Rt_{\a\bb} = \N_{\a} \Ot_{\bb} ,\quad \Rt_{\a\b} =0, $$ $$ \Rt_{ab}
= T_{ab}{}^{\gb} \Ot_{\gb} ,\quad \Rt_{a\bb} =  \N_a \Ot_{\bb}
,\quad \Rt_{a\b} = T_{a\b}{}^{\gb} \Ot_{\gb}. }

Secondly, let us write some expressions useful for later use. We
note that the Bianchi identity $(\N T)_{\a ab}{}^c = 0$,  using
\curcon\  can be written as

\eqn\Ralbcd{R_{\a [ab]c} = \N_{\a} T_{abc} -2 (\g_{c[a})_{\a}{}^\b
R_{b]\b} + (\g_{c})_{\a\b}T_{ab}{}^\b - T_{\a dc} T_{ab}{}^d - T_{\a
[a}{}^d T_{b]d c},} now, we can use the identity

\eqn\idzero{2 R_{\a abc} = R_{\a [ab]c}+ R_{\a [ca]b} -R_{\a [bc]a},
} and the Bianchi identity $(\N H)_{\a abc} =0$ to write \Ralbcd\ as

\eqn\idI{R_{\a abc} = T_{a[b}{}^\b (\g_{c]})_{\b\a} - 2
(\g_{bc})_{\a}{}^\b R_{a\b}.} An identical procedure starting with
$(\N \Tt)_{\ab ab}{}^c =0$ allows us to find

\eqn\idIp{\Rt_{\ab abc}= T_{a[b}{}^{\bb} (\g_{c]})_{\ab\bb} - 2
(\g_{bc})_{\ab}{}^{\bb}\Rt_{a\bb}.} Then, replacing \idI\ and \idIp\
respectively in $(\N T)_{a\a\b}{}^\b = 0$ and $( \N
 T )_{a\ab\bb}{}^{\bb} =0,$ we find

\eqn\idII{\g^b _{\a\b}T_{ba}{}^\b = 8 R_{a\a},\quad \g^b_{\ab\bb}
T_{ba}{}^{\bb} = 8 \Rt_{a\ab}.}

We have enough information to show that the equations \pipidos\ ,
\pipitres\ and \pipicuatro\ are satisfied. From the Bianchi identity
$(\N T)_{\a\b\gb}{}^\b =0$ we obtain

\eqn\BiII{T_{\a\b}{}^d T_{d\gb}{}^\b = 17 R_{\a\gb} + {1\over 4}
R_{\gb\b cd} (\g^{cd})_{\a}{}^{\b}.} Since we need an expression for
$R_{\gb\b cd}$, we can use $(\N T)_{\a\bb a}{}^b=0$, finding

\eqn\BiXXVIII{R_{\gb\b cd} = 2 ( \g_{cd})_{\b}{}^\d \N_{\gb}\O_\d +
T_{c\gb}{}^\e (\g_d)_{\e\b} + T_{c\b}{}^{\eb} (\g_d)_{\eb\gb}.}
Replacing \BiXXVIII\ in \BiII\ ,using the second equation in
\curcon\ , $\N_\a \F= 4\O_\a$ and the constraints coming from
holomorphicity-antiholomorphicity of the BRST current
$T_{a\b}{}^{\gb} = -(\g_a)_{\b\d}P^{\d\gb}$, $T_{a\bb}{}^\g =
(\g_a)_{\bb\db}P^{\g\db}$ we can verify the equation \pipidos.

To verify \pipitres\ and \pipicuatro, we must contract the $a$ and
$b$ indices using $\eta^{ab}$ in \idI\ and \idIp, and use \idII\
together with the relations \dilat.

For deriving the remaining equation of the first set, the coupling to
$\Pi^a \Pb^b$, it is useful to find an expression for $R_{abcd}$,
which can be found from the Bianchi identity $(\N T)_{ab\a}{}^\b$

\eqn\rabcd{R_{abcd} = -{1\over 8} (\g_{cd})_\b{}^\a ( \N_\a
T_{ab}{}^\b  - T_{\a[a}{}^e T_{b]e}{}^\b - T_{\a[a}{}^{\gb}
T_{b]\gb}{}^\b ),} from this equation we construct $\eta^{cd} ( R_{acdb} +
R_{bcda} )$:

\eqn\Rsab{\eta^{cd} ( R_{acdb} + R_{bcda} ) = -{1\over 8 }\eta^{cd}
[(\g_{db})_{\b}{}^\a \N_\a T_{ac}{}^\b + (\g_{da})_{\b}{}^\a \N_\a
T_{bc}{}^\b ] $$ $$ +{1\over 8} \eta^{cd}[ (\g_{db})_{\b}{}^\a T_{\a
[a}{}^e T_{c]e}{}^\b  + (\g_{da})_{\b}{}^\a T_{\a [b}{}^e
T_{c]e}{}^\b ] $$ $$ +{1\over 8} \eta^{cd} [(\g_{db})_{\b}{}^\a
T_{\a [a}{}^{\eb} T_{c]\eb}{}^\b  + (\g_{da})_{\b}{}^\a T_{\a
[b}{}^{\eb} T_{c]\eb}{}^\b]. } Let us consider the right hand side of
\Rsab\ line by line. We can use \idII\ , the Bianchi identity $(\N
R)_{\a a\d \b}{}^\g$ to write

\eqn\BiRaladl{(\g_b)^{\d\a} \N_{\a} R_{a\d} = - 2 \N_a \N_b \F - 2
T_{ab}{}^\g \N_\g \F - 2 (\g_b \g_{ae})^{\d\b}\O_\b R^{e}{}_\d -
(\g_a \g_b)_{\b}{}^\d P^{\b\eb} R_{\eb\d},} and the beta function with
dimension $1$ \pipidos\ to find
the following expression for the first line in the right hand side
of \Rsab\

\eqn\fstRsab{-4 \N_b \N_a \F + 2 T_{ab}{}^C \N_C \F - 4
\eta_{ab}(\g^e)^{\d\b} \O_\b R_{e\d} + 4 (\g_b)^{\d\b}\O_\b R_{a\d}
+ 4 (\g_a)^{\d\b}\O_\b R_{b\d} + {1\over 4} \eta_{ab} \eta^{cd}
T_{c\b}{}^{\db} T_{d\db}^{\b}.} Finding an expression for the second
line is a matter of gamma matrices algebra, once we use \curcon\ .
For this line we find ${1\over 4}\eta_{ab}T_{\b cd} T^{cd\b} -
{3\over 4}T_{c(a}{}^\b T_{b)\b}{}^c $. Using $T_{a\b}{}^{\gb} =
-(\g_a)_{\b\d}P^{\d\gb}$ and some gamma matrices algebra, it is
straightforward to find $T_{\b (a}{}^{\gb} T_{b)\gb}{}^b -{1\over
4}\eta_{ab}\eta^{cd}T_{d\b}{}^{\gb}T_{c\gb}{}^\b$ for the third
line. So, adding the results for the three lines and using \idII\ we
find

\eqn\RsabII{\eta^{cd} ( R_{acdb} + R_{bcda} ) = - 4 \N_b \N_a \F -
T_{c(a}{}^\b T_{b)\b}{}^c + 2 T_{ab}{}^E \N_E \F + T_{\b
(a}{}^{\gb}T_{b)\gb}{}^\b ,} which contains some of the terms in
\pipiuno\ . It is also needed to use $(\N T)_{abc}{}^c=0$ in order
to generate the term $\N^c T_{abc}$. This Bianchi identity gives

\eqn\BiXXVII{\N^c T_{abc} -T_{c[a}{}^e T_{b]e}{}^c -T_{c[a}{}^\e
T_{b]\e}{}^c -\eta^{cd}(R_{acdb}-R_{bcda})=0 .} Finding an
expression for $\eta^{cd}(R_{acdb}-R_{bcda})$ is not difficult
following the description given to compute \RsabII\ . After we
compute it and replace it in \BiXXVII\ we find \eqn\BiXXVIIp{\N^c
T_{abc} +T_{\b [a}{}^{\db} T_{b]\db}{}^\b -2T_{ab}{}^c \N_c \F
+2T_{ab}{}^\g \N_\g \F -2T_{ab}{}^{\gb}\N_{\gb}\F =0.} Combining
\RsabII\ and \BiXXVIIp\ gives the desired beta function equation
\pipiuno.

A similar procedure, but with more steps, is performed to prove the
equations of the second group. To probe \dalPbbb\ one can start by
computing $\{\N_{\ab},\N_{\bb}\}P^{\g\bb} = -\g^c _{\gb\bb}\N_c
P^{\g\bb}+\Rt_{\ab\bb\db}{}^{\bb}P^{\g\db }$. Then we split the
curvature as a scale curvature plus a Lorentz curvature. For the
latter, use $(\N \Tt)_{\ab\bb c}{}^d =0$ to obtain

\eqn\Rabbbcd{\Rt_{\ab\bb cd}(\g^{cd})_{\db}{}^{\bb} = - 180 \N_{\ab}
\Ot_{\db} + (\g^{cd})_{\db}{}^{\bb} \N_{\bb} \Tt_{\ab cd} + 16
\Tt_{\db}{}^{cd} \Tt_{\ab cd} + (\g^{cd} \g^e)_{\db\ab} \Tt_{ecd}, }
so on one hand we will have

\eqn\DDPI{\{\N_{\ab},\N_{\bb}\}P^{\g\bb} = - \N^c T_{c\ab}{}^\g +
\Rt_{\ab\db} P^{\g\db} - 45 \N_{\ab}\Ot_{\db}P^{\g\db} + {1\over 4}
(\g^{cd})_{\db}{}^{\bb} \N_{\bb} \Tt_{\ab cd} P^{\g\db}$$
$$ - 4 \Tt_{\ab cd} \Tt_{\db}{}^{cd} P^{\g\db} + {1\over
4} (\g^{cd}\g^e)_{\db\ab} \Tt_{ecd} P^{\g\db}.} On the other hand,
we can use $\N_{\ab}P^{\b\gb} = \Ct_{\ab}{}^{\gb\b}, \Ct^{\g} = -
P^{\g\db} \Ot_{\db}$ and $\Ct_{cd}{}^\g = 1/10 (\g^a)^{\g\a}
\Rt_{a\a cd}$, which come from antiholomorphicity of the BRST
current, to write

\eqn\DDPII{\{\N_{\ab},\N_{\bb}\}P^{\g\bb} = - 17 \N_{\ab} P^{\g\db}
\Ot_{\db} - 17 P^{\g\db} \N_{\ab} \Ot_{\db} + {1\over 40}
(\g^{a})^{\g\a} \N_{\bb} ( \Rt_{a\a cd}(\g^{cd})_{\ab}{}^{\bb} ).}
Using $(\N \Tt)_{\a bcd}=0$ and $(\Nt H)_{\a bcd}=0$ it is
straightforward to find

\eqn\gRt{(\g^a)^{\g\a}\Rt_{a\a cd} = 10 T_{cd}{}^\g - 10 P^{\g\eb}
\Tt_{\eb cd}.} Since there is a derivative acting on this terms in
\DDPII\ , we make use of $(\N \Tt)_{\bb cd}{}^\g =0$ to find

\eqn\BiTtbbcdg{(\g^{cd})_{\ab}{}^{\bb} \N_{\bb} T_{cd}{}^\g = - 18
\N^d T_{d\ab}{}^\g + (\g^{cd}\g^e)_{\ab\db} \Tt_{ecd} P^{\g\db} + 16
\Tt_{\ab cd} T_{dc}{}^\g.} We can now replace the last two equations
in \DDPII\ and equate it to \DDPI\ . The identity

\eqn\idIII{(\g^{ab})_{(\ab}{}^{\bb}(\g_{ab})_{\gb )}{}^{\db} = - 10
\d_{(\ab}{}^{\bb} \d_{\gb )}{}^{\db} + 8
(\g^a)_{\ab\gb}(\g_a)^{\bb\db},} which can be proved using
$(\g^a)_{(\ab\bb}(\g_a)_{\gb )\db}=0$, will be of help to find
\dalPbbb. A completely analog procedure allows us to arrive to
\Pialdtbb.

To prove \dalPbb\ we make use of the Bianchi identities $(\N R)_{\a
ab\b}{}^\g =0$, $(\N T)_{c\a\b}{}^\g=0$ and the identity
$(\g_a)^{\a\b} R_{\a\b\g}{}^\d = - 2 (\g_a)^{\a\b} R_{\g\a\b}{}^\d$,
which follows from $(\N T)_{\a\b\g}{}^\d =0$, to arrive to

\eqn\idIV{(\g)^{\a\b} ( \N_{\a} R_{ab\b}{}^\g - 2 T_{\a [a}{}^e
 R_{b]e\b}{}^\g - T_{\a[a}{}^{\eb}R_{b]\eb\b}{}^\g) -
8 T_{b}{}^{ac} T_{ac}{}^\g + 8 \N^a T_{ab}{}^\g + 2 T_{ab}{}^\g \N^a
\F
$$ $$ - {1\over 8} (\g)^{\a\b}(\g^{cd})_{\e}{}^\g R_{\a\b
cd} T_{ab}{}^\e + T_{ab}{}^{\eb} (\g^a)^{\a\b}R_{\eb\a\b}{}^\g =0.}
The last term in this equation is zero as can easily seen using $(\N
T)_{\eb\a\b}{}^\g =0$. The first term can be worked out using
\rabcd\ and $(\N T)_{a\eb\b}{}^\g =0$, the curvature in the first
term of the second line can be rewritten using $(\N T)_{\a\b
a}{}^b=0$. The use of $(\N T)_{cdb}{}^\d =0$ will be also needed to
generate \dalPbb. Again, an analog procedure will allow at arrive to
\Piadtbb.

So far, we concentrated on a specific set of beta functions. The
remaining ones can be classified in a third and fourth sets. The
third set involves first order derivatives of the curvatures. We
present it again as the dimension increases.

At dimension $5/2$ we find respectively from the couplings to $J
\Pb^{\bb}$, $\Pi^\a \Jt$, $N^{ac}\Pb^{\bb}$ and  $\Pi^\a \Nn^{bc}$

\eqn\JPbbb{\N^a R_{a\bb} + \N_{(\eb} R_{\bb) \d } P^{\d\eb} + 2
(\N_{\bb}C^{\ab} - R_{\bb\g} P^{\g\ab}) \N_{\ab} \F + 2 C^{\ab}
\N_{\ab} \N_{\bb} \F = 0,}

\eqn\PialJt{\N^b \Rt_{b\a} - \N_{(\d} \Rt_{\a)\eb} P^{\d\eb} + 2
(\N_{\a} \Ct^{\b} + \Rt_{\a\gb} P^{\b\gb})\N_{\b}\F + 2 \Ct^\b \N_\b
\N_\a \F = 0,}

\eqn\NacPbbb{\N^d R_{d\bb ac} + \N_{(\eb} R_{\bb\d) ac} P^{\d\eb} +
2 (\N_{\bb} C_{ac}{}^{\ab} - R_{\bb\g ac} P^{\g\ab}) \N_{\ab} \F + 2
C_{ac}{}^{\db} \N_{\db} \N_{\bb} \F = 0,}

\eqn\PialNtbc{\N^d \Rt_{d\a bc} - \N_{(\d}\Rt_{\a )\eb bc} P^{\d\eb}
+ 2 (\N_{\a} \Ct_{bc}{}^{\g} + \Rt_{\a\db bc} P^{\g\db}) \N_{\g} \F
+ 2 \Ct_{bc}{}^\g \N_\g \N_\a \F = 0.} While at dimension $3$ we
find respectively from the couplings to $J\Pb^b$, $\Pi^a \Jt$,
$N^{ac}\Pb^b$ and $\Pi^a \Nn^{bc}$

\eqn\JPbb{\N^a R_{ab} - T_{ba}{}^c R^{a}{}_c + T_{ba}{}^{\g}
R^{a}{}_{\g} + 3 T_{b\gb}{}^{\a} \N_{\a} C^{\gb} + 2 R_{bc} \N^c \F
+ 2 R_{b\ab} P^{\g\ab}\N_{\g} \F $$ $$ + 2 (\N_b C^{\ab} - R_{b\g}
P^{\g\ab}) \N_{\ab} \F + P^{\d\eb} (\N_{\eb} R_{\d b} + T_{b\d}{}^c
R_{\eb c} + T_{b\eb}{}^{\g} R_{\d\g})=0,}

\eqn\PiaJt{\N^b \Rt_{ba} + T_{ab}{}^{\gb} \Rt^{b}{}_{\gb} + T_{abc}
\Ct^\d T_{\d}{}^{bc} + 3 T_{a\g}{}^{\bb} \N_{\bb} \Ct^{\g} + 2
\Rt_{ab} \N^b \F - 2 \Rt_{a\g} P^{\g\bb} \N_{\bb} \F $$
$$ + 2(\N_a \Ct^{\b} + \Rt_{a\gb} P^{\b\gb}) \N_{\b} \F
- P^{\d\eb} (\N_{\d} \Rt_{\eb a} + T_{\d a}{}^c \Rt_{\eb c} +
T_{a\d}{}^{\gb} \Rt_{\eb\gb})=0,}

\eqn\NacPbb{\N^d R_{dbac} - T_{b}{}^{de} R_{deac} + T_{b}{}^{d\e}
R_{d\e ac} + 3 T_{b\db}{}^{\g}\N_{\g}C_{ac}{}^{\db}+2R_{bdac}\N^d \F
+2R_{b\db ac} P^{\e\db} \N_{\e} \F $$ $$+ 2 (\N_b C_{ac}{}^{\db} -
R_{b\e ac} P^{\e\db}) \N_{\db} \F  +2 R_{b\db ea}C_{c}{}^{e\db} +
P^{\d\eb} (\N_{\eb} R_{\d bac} + T_{b\d}{}^f R_{\eb fac} +
T_{b\eb}{}^{\g} R_{\d\g ac})=0,}

\eqn\PiaNtbc{\N^d \Rt_{dabc} + T_{a}{}^{d\eb} \Rt_{d\eb bc} +
T_{adf} \Ct_{bc}{}^\e T_{\e}{}^{df} + 3 T_{a\d}{}^{\eb}
\N_{\eb}\Ct_{bc}{}^{\d} + 2 \Rt_{adbc} \N^d \F - 2 \Rt_{a\d bc}
P^{\d\eb} \N_{\eb} \F $$ $$ + 2 (\N_a \Ct_{bc}{}^{\d} + \Rt_{a\eb
bc} P^{\d\eb}) \N_{\d} \F + 2 \Rt_{a\d eb} \Ct_{c}{}^{e\d} -
P^{\d\eb} (\N_{\d} \Rt_{\eb abc} + T_{a\d}{}^f  \Rt_{\eb fbc} +
T_{a\d}{}^{\gb} \Rt_{\eb\gb bc})=0.}

The fourth set involves second order derivatives of the
background fields $P^{\a\bb}$, $C_{\a}{}^{\b\gb}$,
$\Ct_{\ab}{}^{\bb\g}$ and $S_{\a\bb}{}^{\g\db}$. There is an
equation at dimension $3$, coming from the coupling to $d_\a
\dt_{\bb}$

\eqn\daldtbb{\N^2 P^{\a\bb} - 2 P^{\g\db} S_{\g\db}{}^{\a\bb} +
T_{de}{}^{\a} T^{de\bb} - 2 \N_{\gb} P^{\d\bb} \N_{\d} P^{\a\gb} - 2
\N_c P^{\a\bb} \N^c \F $$ $$ - 2 (P^{\g\db} \N_{\db} P^{\a\bb} +
P^{\a\db} \N_{\db} P^{\g\bb}) \N_{\g} \F + 2 (P^{\d\gb} \N_{\d}
P^{\a\bb} + P^{\d\bb} \N_{\d} P^{\a\gb}) \N_{\gb} \F =0.} At
dimension $7/2$ we find respectively from the couplings to
$J\dt_{\bb}$, $d_\a \Jt$, $N^{ac}\dt_{\bb}$ and $d_\a \Nn^{bc}$

\eqn\Jdbb{\N^2 C^{\bb} - P^{\a\gb} \N_{[\a} \N_{\gb ]} C^{\bb} -
T_{ac}{}^{\bb} R^{ac} + 2 R_{\g}{}^a \N_a P^{\g\bb} + 2 \N_{\gb}
P^{\a\bb} \N_\a C^{\gb} - C^{\ab} \Rt_{\ab\d\eb}{}^{\bb} P^{\d\eb}$$
$$ + P^{\a\bb} (\N_c R_\a{}^c - \N_{[\d}R_{\gb ]\a} P^{\d\gb}) - 2 (\N_a
C^{\bb} - P^{\g\bb} R_{a\g}) \N^a \F - 2 (P^{\a\bb} \N_\a C^{\gb} +
P^{\a\gb} \N_\a C^{\bb}$$
$$ + P^{\a\bb} R_{\a\g} P^{\g\gb}) \N_{\gb} \F + 2( S P^{\a\bb} + {1\over
4} \St_{cd} (\g^{cd})_{\eb}{}^{\bb} P^{\a\eb} - C^{\gb} \N_{\gb}
P^{\a\bb}) \N_{\a} \F =0,}

\eqn\Jdbb{\N^2 \Ct^{\a} - P^{\b\gb} \N_{[\b}\N_{\gb ]} \Ct^{\a} -
T_{bc}{}^{\a} \Rt^{bc} - 2 \Rt_{\gb}{}^b \N_b P^{\a\gb} - 2 \N_{\gb}
\Ct^\b \N_\b P^{\a\gb} + \Ct^{\b} R_{\b\eb\d}{}^{\a} P^{\d\eb}
$$ $$ - P^{\a\bb} (\N_c \Rt_{\bb}{}^c + \N_{[\d} \Rt_{\gb
]\bb} P^{\d\gb}) - 2 (\N_b \Ct^{\a} + P^{\a\gb} \Rt_{b\gb}) \N^b \F
+ 2 (P^{\a\bb} \N_{\bb} \Ct^{\g} + P^{\g\bb} \N_{\bb} \Ct^{\a}$$
$$ + P^{\a\eb} \Rt_{\eb\gb} P^{\g\gb}) \N_{\g} \F - 2 (S P^{\a\bb} + {1\over
4} S_{cd} (\g^{cd})_{\e}{}^{\a} P^{\e\bb} - \Ct^{\g} \N_{\g}
P^{\a\bb}) \N_{\bb} \F =0,}

\eqn\Nacdbb{ \N^2 C_{ac}{}^{\bb} - P^{\d\eb} \N_{[\d}\N_{\eb
]}C_{ac}{}^{\bb} - R_{deac} T^{de\bb} - 2 R_{d\e ac} \N^d P^{\e\bb}
+ 2 \N_{\db} P^{\e\bb} \N_{\e} C_{ac}{}^{\db} - C_{ac}{}^{\gb}
\Rt_{\gb\d\eb}{}^{\bb} P^{\d\eb} $$ $$ - P^{\b\bb} (\N^d R_{d\b ac}
- \N_{[\d}R_{\eb ]\b ac} P^{\d\eb} + 2 R_{\b\db ea} C_{c}{}^{e\db})
+ 2 \N_{\db} C_{ea}{}^{\bb} C_{c}{}^{e\db} - 2 (\N_d C_{ac}{}^{\bb}
- P^{\e\bb} R_{d\e ac}) \N^d \F
$$ $$ -2 (C_{ac}{}^{\db} \N_{\db} P^{\g\bb}- S_{ac} P^{\g\bb} -
{1\over 4} S_{acbd}(\g^{bd})_{\db}{}^{\bb} P^{\g\db}) \N_\g \F $$ $$
- 2 (P^{\g\bb} \N_\g C_{ac}{}^{\db} + P^{\g\db} \N_\g C_{ac}{}^{\bb}
- P^{\e\bb} R_{\e\g ac} P^{\g\db}) \N_{\db} \F = 0,}

\eqn\daNtbc{ \N^2 \Ct_{bc}{}^{\a} - P^{\d\eb} \N_{[\d}\N_{\eb ]}
\Ct_{bc}{}^{\a} - \Rt_{debc} T^{de\a} + 2 \Rt_{d\eb bc} \N^d
P^{\a\eb} - 2 \N_{\db} \Ct_{bc}{}^\e \N_\e P^{\a\db} +
\Ct_{bc}{}^{\g} R_{\g\eb\d}{}^{\a} P^{\d\eb} $$ $$ - P^{\a\bb} (\N^d
\Rt_{d\bb bc} - \N_{[\d} \Rt_{\eb ]\bb bc} P^{\d\eb} + 2 \Rt_{\bb\d
eb} \Ct_{c}{}^{e\d}) + 2 \N_{\d} \Ct_{eb}{}^{\a} \Ct_{c}{}^{e\d} - 2
(\N_d \Ct_{bc}{}^{\a} + P^{\e\bb} \Rt_{d\bb bc}) \N^d \F
$$ $$ + 2 (\Ct_{bc}{}^{\d} \N_{\d} P^{\a\gb} - \St_{bc} P^{\a\gb}
- {1\over 4} S_{adbc} (\g^{ad})_{\d}{}^{\a} P^{\d\gb})\N_{\gb} \F $$
$$ + 2 (P^{\a\bb} \N_{\bb} \Ct_{bc}{}^{\d} + P^{\d\bb} \N_{\bb}
\Ct_{bc}{}^{\a} + P^{\a\eb} \Rt_{\eb\gb bc} P^{\d\gb}) \N_{\d}
\F=0.}

Finally, at dimension $4$ we find from the couplings to  $J\Jt$,
$J\Nn^{ac}$, $N^{ab}\Jt$ and $N^{ab}\Nn^{cd}$ respectively

\eqn\JJt{\N^2 S - P^{\d\eb} \N_{[\d} \N_{\eb ]} S - R^{ab} \Rt_{ab}
+ 2 \Rt_{a\bb} \N^a C^{\bb} + 2 R_{a\b} \N^a \Ct^\b - 2 \N_{\ab}
\Ct^\b \N_\b C^{\ab}$$ $$ - \Ct^\b (\N^a R_{a\b} - P^{\d\eb}
\N_{[\d}R_{\eb ]\b}) - C^{\bb} (\N^a \Rt_{a\bb} - P^{\d\eb} \N_{[\d}
\Rt_{\eb ]\bb}) + 2 (\Ct^\a R_{b\a} + C^{\ab} \Rt_{b\ab}) \N^b \F
$$
$$ - 2(C^{\ab} \N_{\ab} \Ct^\b
+ P^{\b\ab}(\N_{\ab} S + C^{\gb} \Rt_{\gb\ab} + \Ct^\g R_{\g\ab}))
\N_\b \F - 2 (\Ct^{\a} \N_{\a} C^{\bb} - P^{\a\bb} (\N_\a S $$
$$ + C^{\gb} \Rt_{\gb\a} + \Ct^\g R_{\g\a})) \N_{\bb} \F =0, }

\eqn\JNtac{\N^2 \St_{ac} - P^{\d\eb} \N_{[\d} \N_{\eb ]} \St_{ac} -
R^{ed} \Rt_{edac} + 2 \Rt_{b\db ac} \N^b C^{\db} + 2 R_{b\d} \N^b
\Ct_{ac}{}^\d - 2 \N_{\bb} \Ct_{ac}{}^\d \N_\d C^{\bb} - 2 \N_\d
\St_{ba} \Ct_{c}{}^{b\d}$$ $$ - C^{\bb} (\N^d \Rt_{d\bb ac} -
P^{\d\eb} \N_{[\d} \Rt_{\eb ]\bb} + 2 \Rt_{\bb \d ea}
\Ct_{c}{}^{e\d}) - \Ct_{ac}{}^{\b} (\N^d R_{d\b} - P^{\d\eb} \N_{[\d}
R_{\eb ]\b})$$ $$ +2 (\Ct_{ac}{}^\b R_{d\b} + C^{\bb} \Rt_{d\bb ac})
\N^d \F - 2 C^{\db} \N_{\db} \Ct_{ac}{}^\g \N_\g \F + 4 \St_{ab}
\Ct_{c}{}^{b\g} \N_\g \F - 2 \Ct_{ac}{}^{\b} \N_{\b} C^{\gb} \N_{\gb}
\F =0,}

\eqn\NabJt{\N^2 S_{ab} - P^{\d\eb} \N_{[\d} \N_{\eb ]} S_{ab} -
\Rt^{cd} R_{cdab} + 2 R_{c\d ab} \N^c \Ct^{\d} + 2 \Rt_{c\db} \N^c
C_{ab}{}^{\db} -2 \N_{\gb} \Ct^\d \N_\d C_{ab}{}^{\gb} - 2 \N_{\db}
S_{ca} C_{b}{}^{c\db}$$ $$ - \Ct^{\g} (\N^d R_{d\g ab} - P^{\d\eb}
\N_{[\d} R_{\eb ]\g ab} + 2 R_{\g \db ea} \Ct_{b}{}^{e\db}) -
C_{ab}{}^{\gb} (\N^d \Rt_{d\gb} - P^{\d\eb} \N_{[\d} R_{\eb ]\gb})$$
$$ + 2 (\Ct^\g R_{d\g ab} + C_{ab}{}^{\gb} \Rt_{d\gb ab}) \N^d \F
- 2 C_{ab}{}^{\gb} \N_{\gb} \Ct^\d \N_\d \F + 4 S_{ac}C_{b}{}^{c\db}
\N_{\db} \F - 2 \Ct^{\g} \N_{\g} C_{ab}{}^{\db} \N_{\db} \F =0,}

\eqn\NabNcd{\N^2 S_{abcd} - P^{\d\eb} \N_{[\d} \N_{\eb ]} S_{abcd} -
\Rt^{ef}{}_{cd} R_{efab} + 2 \Rt_{f\eb cd} \N^f C_{ab}{}^{\eb} + 2
R_{f\e ab} \N^f \Ct_{cd}{}^\e - 2 \N_{\eb} \Ct_{cd}{}^\g \N_\g
C_{ab}{}^{\eb}$$
$$ + 2\N_{\eb} S_{afcd} C_{b}{}^{f\eb} + 2 \N_\e
S_{abcd} \Ct_{d}{}^{f\e} - C_{ab}{}^{\eb} (\N^e \Rt_{e\eb cd} -
P^{\d\gb} \N_{[\d} \Rt_{\gb ]\eb cd} + 2 \Rt_{\eb \d ec}
\Ct_{d}{}^{e\d}) - \Ct_{cd}{}^{\e} (\N^e R_{e\e ab}$$
$$- P^{\d\gb} \N_{[\d} R_{\gb ]\e ab}) + 2 (\Ct_{cd}{}^\e R_{e\e ab}
+ C_{ab}{}^{\eb} \Rt_{e\eb cd}) \N^e \F - 2 C_{ab}{}^{\gb} \N_{\gb}
\Ct_{cd}{}^\e \N_\e \F + 4 S_{abcf} \Ct_{d}{}^{f\e} \N_\e \F $$ $$ - 2
\Ct_{cd}{}^{\g} \N_{\g} C_{ab}{}^{\eb} \N_{\eb} \F + 4
S_{afcd}C_{b}{}^{f\eb}\N_{eb}\F=0.}

Since the Bianchi identities allow to write the curvature components
in terms of the torsion components, we expect that the beta
functions of the third set will be implied by the eight beta
functions already proven, i.e first and second set. In the same way
we expect that the beta functions of the fourth set will also be
implied by the first two sets of beta functions since the
constraints coming from holomorphicity and antiholomorphicity of the
BRST current allows to relate the background fields to some
components of the torsion. This is not too hard to check in the case
of lower dimension, for example, at dimension $5/2$ consider the
beta functions coming from the coupling to $J\Pb^{\bb}$

\eqn\JPbbb{\N^a R_{a\bb} + \N_{(\eb} R_{\bb )\d} P^{\d\eb} + 2
(\N_{\bb} C^{\ab} - R_{\bb\g} P^{\g\ab}) \N_{\ab} \F + 2 C^{\ab}
\N_{\ab} \N_{\bb} \F = 0.} By using $R_{a\bb} = T_{a\bb}{}^\g \O_\g$
and $R_{\bb\d} = \N_{\bb}\O_\d$, which follow from the definition of
the curvature, and $C^{\bb} = P^{\a\bb}\O_\a$, which follows from
the antiholomorphicity constraints, we find that \JPbbb\ can be
written as

\eqn\last{(\N^c T_{c\bb}{}^{\a} -2 \N_{\bb} P^{\a\gb} \N_{\gb} \F +
2 P^{\a\gb} \N_{\gb} \N_{\bb} \F )\O_\a =0,} so, the beta function
\dalPbbb\ with dimension $2$ implies \JPbbb\ . Similarly we
checked that \Pialdtbb\ implies \PialJt\ and that the beta functions
with dimension $5/2$ \dalPbb\ and \Piadtbb\ imply respectively the
beta functions with dimension $3$ \JPbb\ and \PiaJt\ .

We have not found proofs for the vanishing of the remaining beta
functions, since this task becomes clumsy as the dimension increases.
Nevertheless, given the above explanation,  we consider our work
sufficient to assure that the beta functions vanish as a consequence of the
classical BRST symmetry of the action for the Type II superstring in
a generic supergravity background.

\vskip 15pt {\bf Acknowledgements:} We would like to thank Nathan
Berkovits, Brenno Carlini Vallilo and  Daniel L. Nedel. The work of OC is supported
Fundaci\'on Andes, FONDECYT grant 1061050 and Proyecto Interno
27-05/R from UNAB. The work of O.B is supported by CAPES, grant 33015015001P7.

\listrefs

\end